# Layer thickness and substrate effects on superconductivity in epitaxial FeSe films on BLG/SiC(0001)

Yongsong Wang[1,⊥], Haojie Guo*,[1,⊥], Ane Etxebarria[2], Sandra Sajan[1], Sara Barja[1,2,3,4], and Miguel M. Ugeda*,[1,3,4]

[1]*Donostia International Physics Center, Paseo Manuel de Lardizábal 4, 20018 San Sebastián, Spain.*

[2]*Department of Polymers and Advanced Materials, University of the Basque Country (UPV/EHU), 20018 San Sebastián, Spain.*

[3]*Centro de Física de Materiales, Paseo Manuel de Lardizábal 5, 20018 San Sebastián, Spain.*

[4]*Ikerbasque, Basque Foundation for Science, 48013 Bilbao, Spain.*

[⊥]*These authors contributed equally to this work.*

*Corresponding authors: haojie.guo@dipc.org and mmugeda@dipc.org*

**Abstract:** *The layered nature and simple structure of FeSe reveal this iron-based superconductor as a unique building block for the design of artificial heterostructure materials. While superconductivity develops in ultrathin films of FeSe on $SrTiO_3$ substrates, it remains unclear whether it can be developed on more chemically inert, layered materials such as graphene. Here, we report on the characterization of the structural, chemical and electronic properties of few-layer FeSe on bilayer graphene grown on SiC(0001) using low-temperature scanning tunneling microscopy/spectroscopy (STM/STS) and X-ray photoelectron spectroscopy (XPS). STM imaging of our FeSe films with thicknesses up to three layers exhibit the tetragonal crystal structure of bulk FeSe, which is supported by XPS spectra consistent with the FeSe bulk counterpart. While our STS measurements at 340 mK reveal a metallic character for few-layer FeSe on BLG/SiC(0001), they show an absence of superconductivity, as the low-lying electronic structure exhibits a spatially anisotropic dip-like feature robust against magnetic fields. Superconductivity in FeSe/BLG/SiC(0001), however, emerges for thicker films with a transition temperature of 6 K. Our results underscore the significance of the substrate as key factor driving the suppression superconductivity in FeSe in the 2D limit.*

# 1. Introduction

Iron-based materials exhibit a broad range of intriguing novel physical phenomena such as unconventional superconductivity, quantum criticality, and topological properties [1-7]. Among these materials FeSe stands out due to its simple crystal structure, which shows tetragonal symmetry (see Figure 1a-b) belonging to the P4/nmm space group, and its unique status as a layered material which has been possible to synthesize down to the monolayer level. This structural flexibility not only enables detailed investigation of its fundamental properties at the two-dimensional (2D) limit, but also makes FeSe an ideal candidate for use as a building block for the development of novel quantum materials in artificial heterostructures [4,5,8-20]. FeSe is mostly known for hosting superconductivity with marked layer-dependent properties. While bulk FeSe is a superconductor with a critical temperature ($T_C$) of ≈ 8 K [21], single- and few-layer FeSe have shown the highest ever reported $T_C$ within the iron-based superconductors family, reaching up to 109 K when grown on strontium titanate ($SrTiO_3$) [16,22-48]. This enhancement of $T_C$ has been attributed to interfacial coupling between FeSe and $SrTiO_3$ through different mechanisms, including increased electron-phonon coupling and charge transfer effects [24,27-29,33,39,49]. Therefore, the large sensitivity of the superconducting properties of ultrathin films of FeSe shown on the $SrTiO_3$ has raised the interest in achieving the epitaxial growth them in van der Waals materials for diverse purposes.

Graphene, the most versatile among the 2D materials, is also an outstanding template for the epitaxial growth of various families of 2D materials since it preserves the electronic structure [50-52] and collective electronic phases such as superconductivity [53] of the as-grown 2D material. Regarding FeSe, while scarce previous efforts have shown that the epitaxial growth of ultrathin films of FeSe is plausible on graphene, puzzling results were found concerning their superconducting behavior [54-56]. For instance, while superconductivity was found to be absent in monolayers of FeSe grown on bilayer graphene (BLG) [54], superconductivity has been detected in monolayers of FeSe on trilayer graphene (TLG) with a superconducting gap size of 1 meV [55,56]. More intriguingly, bilayer and few-layers (up to 8 layers) FeSe do now exhibit superconductivity on BLG, with a SC gap size of approximately 2 meV [54].

In this work, we probe the existence superconductivity in the ultimate 2D limit of FeSe, from monolayer to trilayer, on epitaxial bilayer graphene substrates by combining a comprehensive structural, chemical, and electronic characterization using low-temperature (340 mK) scanning tunneling microscopy/spectroscopy (STM/STS) and X-ray photoelectron spectroscopy (XPS). Our experimental STM/STS data indicate that the as-grown FeSe layers exhibit a tetragonal crystal structure and display metallic characteristics in their electronic spectrum, similar to its superconducting bulk counterpart. However, our high-resolution STS data near $E_F$ reveals a spatially anisotropic V-shape dip featuring a strong resilience against out-of-plane magnetic field of up to 11 T. This result allows us to preclude existence of superconductivity in this material for up to three layers down to 340 mK. In contrast, superconductivity is present in multilayers (on average > 20 layers) FeSe grown on BLG with a $T_C$ close to 6 K.

## 2. Results and discussion

Few-layer FeSe was grown by molecular beam epitaxy (MBE) on high-quality BLG formed through the graphitization of the surface of SiC(0001) (see Methods section). This gives rise to the formation of the vertical heterostructure FeSe/BLG/SiC(0001), which is schematically represented in Figure 1b. The surface morphology of the FeSe films is shown in the STM topography of Figure 1c, where exposed regions of BLG/SiC(0001) coexist with FeSe areas of varying thickness. From the apparent height profile recorded along the blue arrow direction, and depicted in Figure 1d, it is inferred an interlayer separation between successive FeSe of 5.8 Å, in agreement with the *c*-axis unit cell size of bulk FeSe [21]. Consequently, the surface is predominantly covered by monolayer FeSe, which grows continuously across the atomic steps of BLG/SiC with full rotational disorder over the graphene. In addition to the complete monolayer, large regions of bilayer and trilayer FeSe were present, the latter as large as $100 \times 100$ nm$^2$.

Representative atomically resolved STM topography images of the BLG/SiC(0001) and monolayer FeSe surfaces are shown in Figure 1e-f, respectively, revealing distinctive atomic arrangements. The BLG/SiC(0001) surface is characterized by a threefold symmetric atomic registry, as expected from the honeycomb structure of graphene, along with a superimposed moiré-like superstructure arising from the reconstruction of SiC(0001) surface beneath [57]. In contrast, the surface of the monolayer FeSe does not exhibit any long-range superstructures and features a twofold square atomic arrangement, with an average separation between the topmost atoms of Se-Se of $3.7 \pm 0.2$ Å. This value agrees with the in-plane lattice constant of bulk FeSe [21,58,59], as well as those for FeSe in the monolayer limit [54-56]. Additionally, the observed square-like structure is consistent with the tetragonal crystal structure of FeSe, as reported in its bulk state [21,59]. It is noteworthy that the atomic registry of the bilayer and trilayer FeSe exhibits the same twofold square symmetry and geometric characteristics as observed in the monolayer.

Next, we probe the chemical properties of few-layer FeSe by means of XPS measurements at room temperature using the same samples previously employed for the STM/STS measurements. Figure 2a shows a survey scan XPS spectrum acquired on FeSe/BLG/SiC(0001). The main core level peaks of the different elements are highlighted in black, while LMM Auger transitions are indicated in green. As expected, carbon, silicon, iron and selenium are the dominant chemical species resolved in the spectrum, along with a faint and minimal peak related to oxygen, which we associate with contaminants arising from the transfer procedure between the STM and XPS chambers (see Methods section) and also inherent gas background from the XPS chamber (mainly H$_2$O). This corroborates the cleanliness of our MBE-grown samples of FeSe on BLG/SiC(0001), ruling out possible contaminants embedded in the FeSe layers. To further characterize the chemical state of Fe and Se, we measured the binding energies of the core level of Fe 2p and Se 3d with higher resolution. The spectra of both core level peaks are shown in Figure 2b-c, respectively, where the peaks are split due to spin-orbit coupling. The main peak of Fe 2p$_{3/2}$ lies at a binding energy of 707 eV, and that of Se 3d$_{5/2}$ at a binding energy of 54.1 eV. These values are in good agreement with those obtained for bulk tetragonal FeSe [60,61], indicating that our few-layer FeSe also hosts the same chemical bonding structure.

Once the structural and chemical properties of few-layer FeSe are clarified, and shown to match those of the superconducting bulk FeSe crystals, we now focus on its electronic properties in the few-layer limit using STS measurements at 4.2 K. Figure 3a shows typical d$I$/d$V$ spectra measured on one to three layers of FeSe over a large bias range. In all cases, the large-scale electronic structure shows the same main features, i.e., a broad peak at a bias voltage $V_S$ = -0.28 V for occupied states (indicated by a blue arrow) and a V-shape local density of states (LDOS) with a non-zero minimum at the Fermi level ($E_F$). The broad peak, previously observed in few-layer FeSe [54], is tentatively attributed to the Fe-derived band of $d_{zx}$ and $d_{zy}$ orbital character and small dispersion along Γ-M calculated for the isolated monolayer of FeSe [19,62-64]. This suggests a weak interaction of FeSe with the graphene underneath. A close-up, high-resolution d$I$/d$V$ spectra acquired at 4.2 K around $E_F$ is shown in Figure 3b. As seen, the typical width of the V-shape dip feature is 25 ± 10 mV. Although this feature is, in principle, compatible with the existence of superconductivity, the significant spatial fluctuations in the shape and width of the dip feature at the nanometer scale (Figure 3c, Figure S1 and Figure S2), despite the high crystallinity of the FeSe layers, suggest a different origin. Indeed, similar dip features unrelated to superconductivity have been previously observed in monolayers of transition metal dichalcogenides on graphene substrates [53,65,66], whose origin has been attributed to Coulomb blockade effect in the STM tunneling junction [67] or phonon-assisted inelastic tunneling [68].

To better probe the existence of superconductivity in these films, we carried out high-resolution STS measurements at 340 mK, the base temperature of our STM instrument. Figure 4 shows a typical high-resolution d$I$/d$V$ spectra acquired from monolayer to trilayer FeSe, where we observe a second dip around $E_F$ with a spatially dependent width in the range of 1-3 meV (Figure S3). This behavior closely resembles to that observed in STS data recorded at higher temperatures and discussed above (note the difference in the energy window). To unambiguously evaluate the existence of superconductivity in the FeSe layers, we characterized the evolution of this dip feature under an out-of-plane magnetic field ($B_\perp$). To minimize the impact of the inherent spatial variations of the dip structure, we performed spatially averaged STS measurements over a 3×3 nm$^2$ area. Figure 4b-d shows the results obtained for monolayer, bilayer and trilayer FeSe, respectively. We observe that the dip structure — its shape, depth, and full width at half maximum — remains largely unaffected by the magnetic field up to 11 T. These results allow us to rule out the existence of a superconducting ground state in few-layer FeSe on BLG/SiC(0001) down to 340 mK.

Lastly, we carried out similar STS measurements on multilayer FeSe (on average > 20 layers) on BLG/SiC(0001) (see Methods section). In these samples, FeSe retains the same atomic configuration and a similar electronic structure to that in the ultrathin layer limit FeSe (see Figure S4a-d in the SI). However, our high-resolution d$I$/d$V$ spectra acquired on multilayer FeSe now show a robust superconducting gap around $E_F$, as shown in Figure 5a (see also Figure S5). This superconducting gap remains visible in our STM d$I$/d$V$ spectra at temperatures as high as 4.2 K (see Figure S4e,f), and shows a $T_C$ = 6 K (Figure S6). The width of this gap, measured as the distance between the coherence peaks position, is 2 ± 0.2 meV, which is comparable to those reported for bulk crystals [69-71] and multilayers [23,25,32]. From the characteristics of the quasiparticle tunneling gap, we infer that superconductivity in multilayer FeSe falls beyond the conventional $s$-wave BCS theory description, thus being compatible recently proposed symmetries such as nodal superconductivity [23,32,69-71]. To confirm that this gap has a superconducting origin, in Figure 5b we show the evolution of this gap under the presence of out-of-plane magnetic field ($B_⊥$), from which we observe a gradual weakening of the gap with the applied field. Finally, the superconducting gap in multilayer FeSe exhibits strong robustness and spatial homogeneity, as evidenced by the d$I$/d$V(r,V)$ map depicted in Figure 5c. All these results highlight the critical influence of FeSe thickness and the substrate on the emergence of the superconducting phase.

The results presented in this work provide valuable new insight into the electronic properties and development of superconductivity of ultrathin FeSe films on graphene substrates. Our high-resolution STS data, acquired after a stringent structural and chemical analysis of our films, are consistent with reports of the absence of superconductivity in monolayer FeSe [54,55], but are in stark contrast to its observation in the case of bilayer and trilayer FeSe. In the latter cases, the reported quasiparticle tunneling spectra show a V-shape dip, with widths of 2 meV and 0.6 meV, respectively, which has been reported as the superconducting gap with a $T_C$ close to 3 K [54,55]. While our STS measurements at 340 mK also unveil a V-shape dip in the electronic structure at $E_F$ on single-to-three layers of FeSe, we unambiguously rule out superconductivity as the origin of this dip feature. Since magnetic field break time reversal symmetry and thus hinders the Cooper pair formation, our $B_⊥$-dependent STS data provide solid evidence for the absence of superconductivity in FeSe in the 2D limit on graphene.

## 3. Conclusions

In summary, we have grown epitaxial few-layer FeSe on BLG/SiC(0001) to investigate their structural, chemical, and electronic properties *via* XPS and low-temperature (340 mK) STM/STS with an electronic temperature resolution of ~ 80 μeV. Our STM and XPS data demonstrate the high-quality synthesis from the monolayer up to three layers of FeSe on BLG/SiC(0001), which display a tetragonal atomic structure as in bulk FeSe. In particular, the XPS experiments rule out the existence of contaminants embedded in the crystal structure that could affect superconductivity. Our subsequent high-resolution STS measurements provide evidence for the suppression of superconductivity in 1-3 layers of FeSe down to 340 mK, as evidenced from the static quasiparticle spectra at $E_F$ with magnetic field. Our work provides accurate information regarding structural quality of epitaxial ultrathin FeSe films as well as the development of superconductivity in the ultimate 2D limit, in particular when becomes part of a graphene-based vertical heterostructure.

## 4. Methods

**Growth of FeSe on BLG/SiC(0001)**

Few-layer FeSe were fabricated on the surfaces of BLG terminated n-type doped 6H-SiC(0001) in a custom-built ultra-high vacuum (UHV) based MBE system. The growth starts by carrying out a slow and throughout outgassing of the SiC(0001) wafer at 800 ºC and continued by annealing at 1400 ºC for 35 min to graphitize the surface. High quality iron and selenium (MaTeck, 99.999 %) were employed as sources of Fe and Se, which were sublimated from an e-beam evaporator and a Knudsen cell, respectively. Both chemical elements were deposited simultaneously on BLG/SiC(0001) while keeping a constant substrate temperature of 450 ºC. A typical growth time of 60 minutes were necessary to achieve a nominal surface coverage as that shown in Figure 1c (nominal coverage ~2 ML), which encompass the coexistence of regions with one and up to three layers of FeSe. To obtain thicker films of FeSe as that shown in Figure S4 in the Supporting Information, we followed the same procedure except for a lower substrate temperature (350 ºC). After the growth, we capped the FeSe layers with an amorphous thick layer of Se deposited at room temperature, which prevents oxidation during its transport in air between the MBE and the STM chambers. The Se capping layers were eliminated prior to the STM measurements by annealing at 300 ºC for 30 min without compromising the pristine quality of FeSe/BLG/SiC(0001).

**XPS measurements**

XPS data were collected in a Phoibos 150 NAP (Specs GmbH) analyzer using a μFOCUS 600 X-ray monochromator (Specs GmbH), with an Al Kα anode (1486.7 eV) and a spot size of 300 μm. The FeSe/BLG/SiC(0001) sample was directly transferred from the STM chamber (USM1800) to the XPS chamber in a UHV suitcase, without exposing it to air. The survey spectrum was collected at 30 eV pass energy, while the core levels of interest were collected at 20 eV. During measurements, the sample was at room temperature, and the system was under UHV conditions. Data was analyzed using CasaXPS (Casa Software Ltd, Teignmouth, UK) software. The binding energy scale is calibrated to the Fermi edge.

**STM/STS measurements and tip calibration**

STM/STS experiments were conducted in two independent UHV chambers, each housing a commercial STM (UNISOKU Co., Ltd) that operates under different temperature conditions. The first system, a cryogen-free closed-cycle STM (USM1800), enables stable measurements at a base temperature of 3.6 K. The second STM (USM1300) allows experiments at temperatures of either 4.2 K or 340 mK and is equipped with a superconducting magnet capable of providing a magnetic field up to 11 T, normal to the sample surface. The specific temperatures at which data were collected are explicitly noted in the figure captions. STS data were acquired using standard lock-in technique, where an AC modulation voltage ($V_{mod}$) was applied to the STM bias voltage during data acquisition, with modulation frequencies of $f = 911$ Hz (USM1800) or $f = 833$ Hz (USM1300). Prior to each experimental run, the STM tip (made of Pt/Ir) was metallized on single crystals of Au(111) or Cu(111) and calibrated against their respective Shockley surface states. Special attention was given to ensuring a flat density of states of the tip near the Fermi level. All the acquired STM/STS data were processed and analyzed using the freeware WSxM [72].

*Acknowledgements*

M.M.U. acknowledges support by the European Union ERC Starting grant LINKSPM (Grant #758558) and by the grant PID2020-116619GB-C21 funded by MCIN/AEI/10.13039/501100011033. S.B. acknowledges support by the European Union ERC Starting grant COSAS (Grant #101040193). H.G. acknowledges funding from the EU NextGenerationEU/PRTR-C17.I1, as well as by the IKUR Strategy under the collaboration agreement between Ikerbasque Foundation and DIPC on behalf of the Department of Education of the Basque Government. S.S. acknowledges enrollment in the doctorate program "Physics of Nanostructures and Advanced Materials" from the "Advanced polymers and materials, physics, chemistry and technology" department of the Universidad del País Vasco (UPV/EHU).

*Author contributions*

M.M.U. conceived the initial project. H.G. and M.M.U. designed the research strategy. Y.S. grew the samples with inputs from H.G. and M.M.U.. Y.S. and H.G. measured the STM/STS data with the help of S.S. and M.M.U.. H.G. and Y.S. analyzed the STM/STS data. A.E. measured and analyzed the XPS data. H.G. wrote the manuscript with inputs from A.E. and M.M.U.. H.G. and M.M.U. coordinated this project. S.B. has participated in the coordination of the STM and XPS experiments. All authors contributed to the scientific discussion and manuscript revisions.

*Data availability*

The data that support the findings of this study are available from the corresponding authors upon reasonable request.

*Competing interests*

The authors declare no competing interests.

# Figures

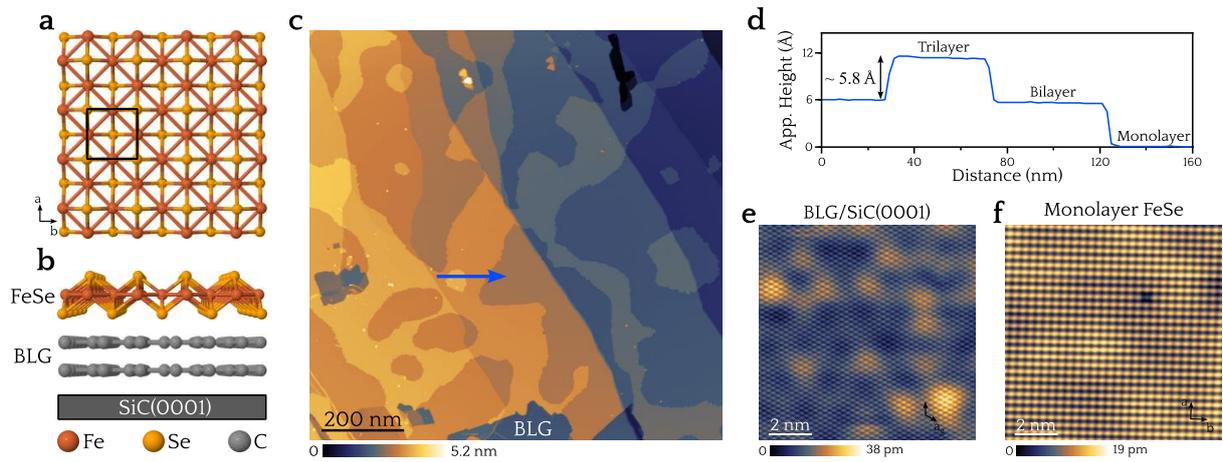

**Figure 1. Structural properties of few-layer FeSe grown on BLG/SiC(0001).** a) Surface atomic arrangement of monolayer FeSe. The black square indicates the surface unit cell, which has a periodicity of 3.9 Å. b) Side view of the tetragonal crystal structure of monolayer FeSe grown on the surfaces of BLG/SiC(0001). c) Large-scale STM topography image revealing the surface morphology of FeSe/BLG/SiC(0001) samples grown by MBE. d) Apparent height profile measured along the direction marked by the blue arrow in (c), passing through regions of trilayer, bilayer and monolayer FeSe. e,f) Atomically resolved STM images showing the different atomic registry of the surfaces of BLG/SiC(0001) and monolayer FeSe, respectively. Acquisition parameters: c) $V_{set} = 1$ V, $I_{set} = 20$ pA, T = 3.6 K. e) $V_{set} = -1$ V, $I_{set} = 0.1$ nA, T = 3.6 K. f) $V_{set} = -1$ V, $I_{set} = 0.1$ nA, T = 3.6 K.

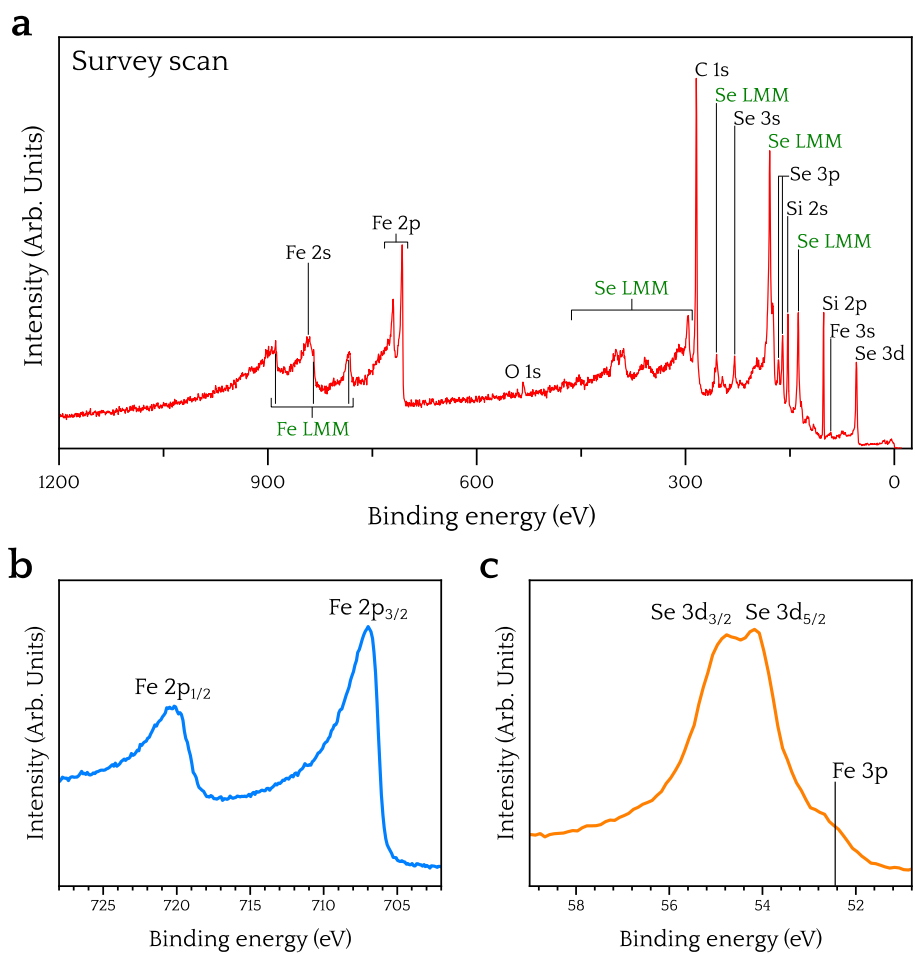

**Figure 2. Chemical properties of few-layer FeSe.** a) XPS survey scan spectrum. The main peaks are identified as the different core levels of Fe, Se, C, Si, and with a residual amount of O. LMM Auger transitions of the elements are highlighted in green as contrast. b,c) High resolution XPS spectra probing the Fe 2p and Se 3d core level, respectively. Both core levels show a splitting due to spin-orbit coupling effect.

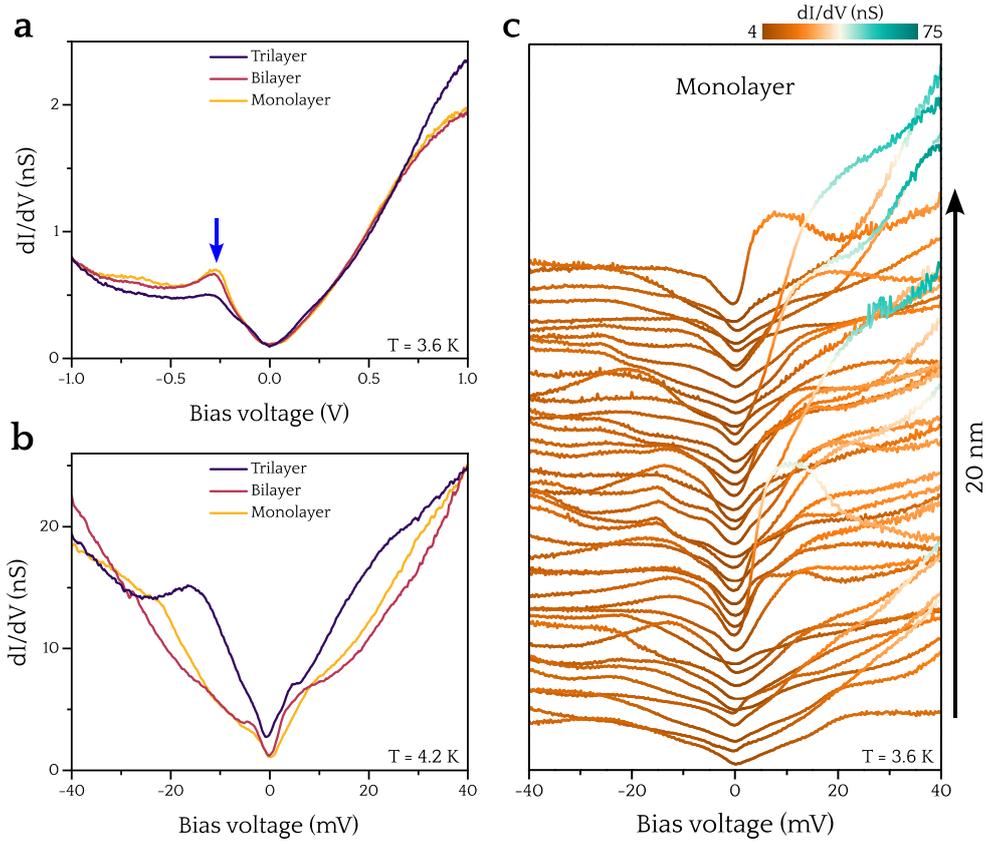

**Figure 3. Electronic structure of few-layer FeSe.** a) Large-scale d$I$/d$V$ curves showing the electronic structure of FeSe for different layer thicknesses. b) Higher resolution d$I$/d$V$ curves acquired near the $E_F$ ($V_S$ = 0 V). c) Waterfall plot of spatially resolved high-resolution d$I$/d$V$ spectra measured around $E_F$ in monolayer FeSe along a linear path of 20 nm. Acquisition parameters: a) $V_{mod}$ = 3.5 mV, $T$ = 3.6 K. b) $V_{mod}$ = 200 μV, $T$ = 4.2 K. c) $V_{mod}$ = 250 μV, $T$ = 3.6 K.

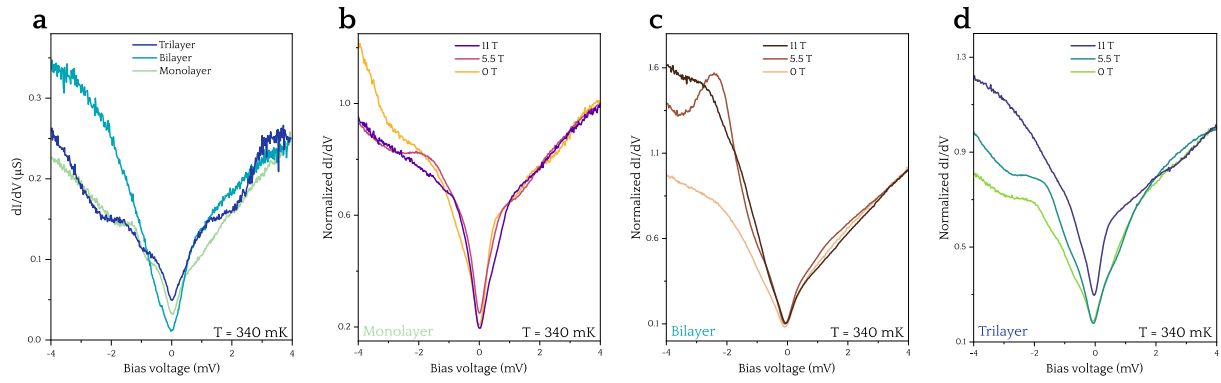

**Figure 4. Absence of superconductivity in few-layer FeSe down to 340 mK.** a) Set of high-resolution dI/dV spectra displaying the dip feature for different layer thickness of FeSe. c) Out-of-plane magnetic field dependent dI/dV spectra showing the resilience of the dip feature in monolayer, bilayer, and trilayer FeSe, respectively. All curves are spatially averaged over regions of 3×3 nm$^2$. Acquisition parameters: a) $V_{mod}$ = 20 µV, T = 340 mK. b-d) $V_{mod}$ = 30 µV, T = 340 mK.

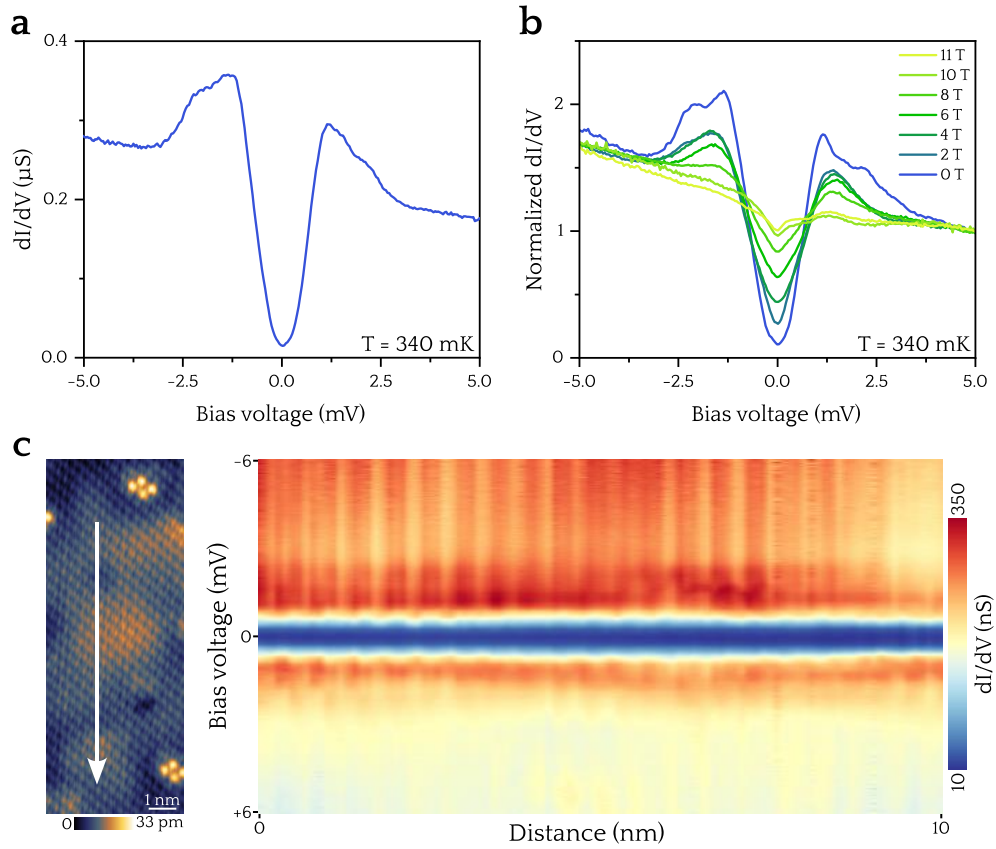

**Figure 5. Emergence of robust superconductivity in multilayer FeSe.** a) High-resolution d$I$/d$V$ spectra displaying the typical tunneling superconducting gap structure recorded on multilayer FeSe. b) Sequentially measured d$I$/d$V$ spectra as function of the out-of-plane magnetic field. c) Spatially resolved d$I$/d$V$ spectra recorded along the white arrow indicated in the topography image on the left. Acquisition parameters: a) $V_{mod}$ = 50 µV, $T$ = 340 mK. b) $V_{mod}$ = 50 µV, $T$ = 340 mK. c) $V_{set}$ = 100 mV, $I_{set}$ = 50 pA, $T$ = 340 mK (topography image) and $V_{mod}$ = 50 µV, $T$ = 340 mK (STS).

# Supplementary information for

# Layer thickness and substrate effects on superconductivity in epitaxial FeSe films on BLG/SiC(0001)

Yongsong Wang, Haojie Guo*, Ane Etxebarria, Sandra Sajan, Sara Barja, and Miguel M. Ugeda*

*Corresponding authors: haojie.guo@dipc.org and mmugeda@dipc.org

**This PDF includes:**

**Supplementary note 1: Spatially varying dip feature near Fermi level at 4.2 K**

**Supplementary note 2: Spatial mapping of the dip structure variation at 4.2 K**

**Supplementary note 3: Spatially varying dip feature near Fermi level at 340 mK**

**Supplementary note 4: Emergence of superconductivity in multilayer FeSe**

**Supplementary note 5: Electronic structure near Fermi level in multilayer FeSe**

**Supplementary note 6: Temperature dependence evolution of the superconducting gap in multilayer FeSe**

**Supplementary note 1: Spatially varying dip feature near Fermi level at 4.2 K**

In the main text, we discussed the presence of a dip feature in the electronic spectrum of few-layer (1-3) FeSe near the Fermi level. This dip deviates from the expected characteristics of a coherent superconducting gap due to its variable size and pronounced spatial fluctuations, even when measured at different points on the surface using the same tip apex. This behavior, illustrated in Figure S1, remains consistent across different FeSe layer thicknesses (1 to 3 layers). The origin of this dip is unknown to us at the present time, and further experiments are desirable to assess its nature.

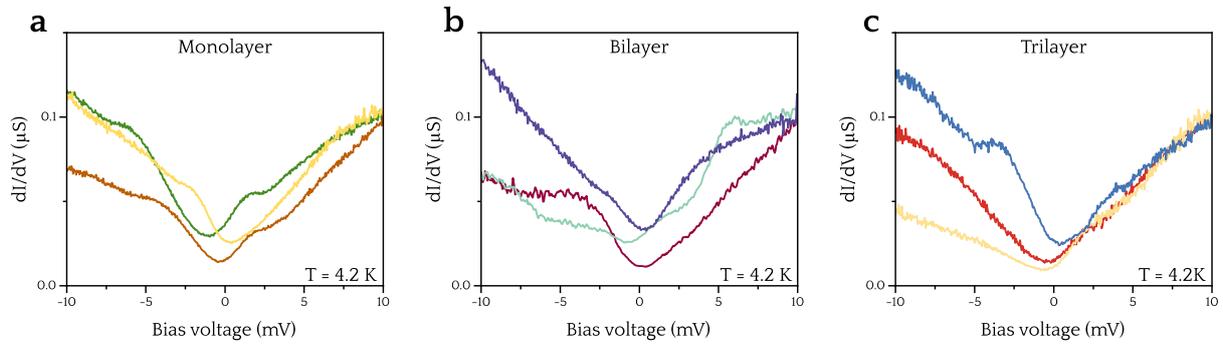

**Figure S1. Anisotropic dip structure near Fermi level at 4.2 K.** a-c) Set of different d$I$/d$V$ spectra recorded on randomized points of the surfaces of few-layer FeSe, displaying a large variability in the electronic structure near Fermi level. Acquisition parameters: a-c) $V_{mod}$ = 200 µV, $T$ = 4.2 K.

**Supplementary note 2: Spatial mapping of the dip structure variation at 4.2 K**

In addition to the single-point spectra presented in Figure S1, where fluctuations in the dip feature are noticeable, a more direct method to visualize and confirm such variations across the surfaces of few-layer FeSe is by acquiring consecutive d$I$/d$V$ spectra along a defined spatial extent. The top panels of Figure S2 illustrate these data for one and up to three layers of FeSe, covering a spatial length of 20 nm in each case. We also show the average curve of each data set in the bottom panels of Figure S2.

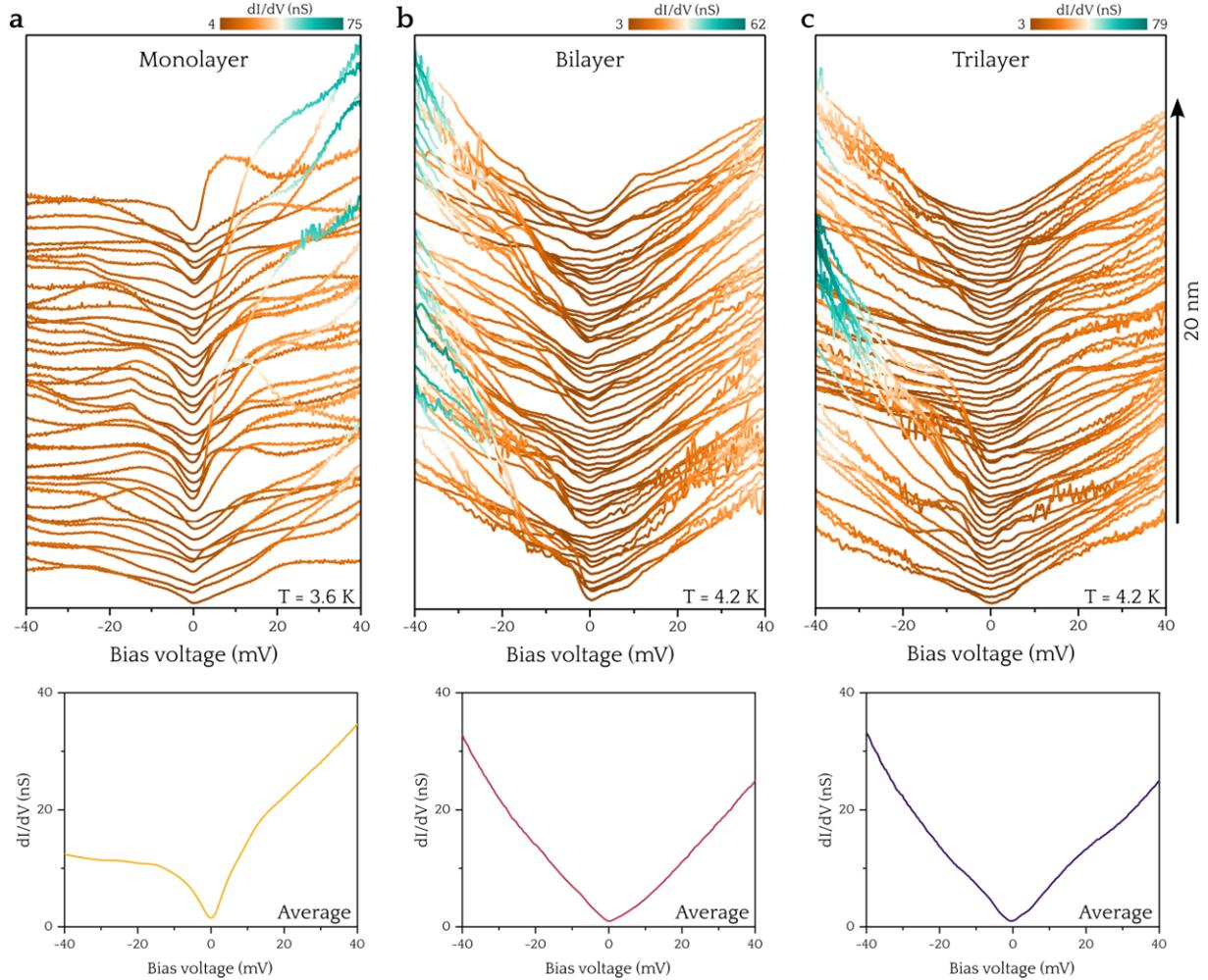

**Figure S2. Spatial mapping of the dip structure variation at 4.2 K.** a-c) Evolution of the electronic structure near the Fermi level along a spatial extension of 20 nm recorded on the surfaces of monolayer, bilayer and trilayer FeSe, respectively (top panels). In the bottom panels, we show the corresponding spatially averaged d$I$/d$V$ curve of each data set. Acquisition parameters: a) $V_{mod}$ = 250 μV, $T$ = 4.2 K. b,c) $V_{mod}$ = 200 μV, $T$ = 4.2 K.

**Supplementary note 3: Spatially varying dip feature near Fermi level at 340 mK**

We conducted STS measurements at 340 mK on few-layer FeSe on BLG/SiC(0001) to probe the existence of superconductivity at this temperature. In the recorded STS spectra, we observed a dip structure similar to that measured at higher temperatures (note the differences in the energy window). In a similar way, we argue that this dip does not correspond to a coherent superconducting gap due to the variable size of the supposed SC gap and the spatial fluctuation observed when spectra are taken at different locations on the sample. All these behaviors are illustrated in Figure S3 for the three different layer thicknesses of FeSe grown in this work. More importantly, we further proved that this dip-structure is not related with superconductivity through its behavior under the presence of an external magnetic field (see Figure 4 in the main text).

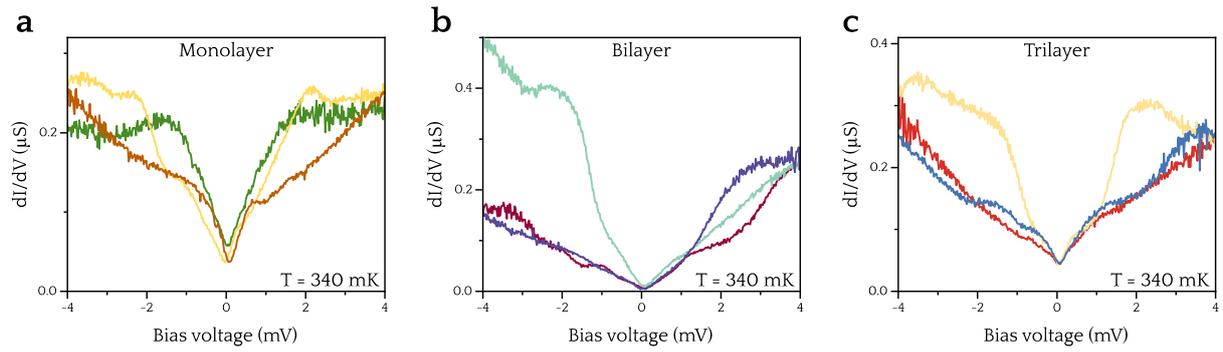

**Figure S3. Anisotropic dip structure near Fermi level at 340 mK.** Differential conductance spectra recorded at different points on the surfaces of monolayer, bilayer and trilayer FeSe, revealing fluctuation of the dip structure around Fermi level. Acquisition parameters: a-c) $V_{mod}$ = 20 μV, $T$ = 340 mK.

**Supplementary note 4: Emergence of superconductivity in multilayer FeSe**

To investigate the impact of the layer thickness of FeSe in the development of superconductivity in this material, we grew thin films of FeSe on BLG/SiC(0001) using a lower growth temperature (see Methods section in the main text). This growth condition resulted in the formation of 3D islands of FeSe with a thickness on average > 20 layers on the surfaces of BLG/SiC(0001), as shown for instance in Figure S4a,b. A typical atomically resolved STM image of the thin film FeSe is displayed in Figure S4c, where the atomic arrangement exhibits a twofold square symmetry, consistent with few-layer FeSe. This indicates that it hosts the very same tetragonal crystal structure. Additionally, the broad electronic structure of the multilayer (Figure S4d) closely matches that of few-layer FeSe, as evidenced from the broad peak located at -0.28 V and the V-shape dip at $E_F$. However, high-resolution STS data recorded near the $E_F$, as that shown in Figure S4e, reveal now the existence of a coherent superconducting gap in the differential conductance spectra already at 4.2 K, with well-defined superconducting coherence peaks. Furthermore, the SC gap shows a rather more homogeneous behavior over the surface, as illustrated in Figure S4f, contrasting with the spatial fluctuations observed for the dip structure of few-layer FeSe. This further supports that it originates from the existence of a superconducting state in multilayer FeSe (see also Figure 5 in the main text). Our results suggest that the reduction of the dimensionality could be the major driving force of the suppression of superconductivity in FeSe in the 2D limit.

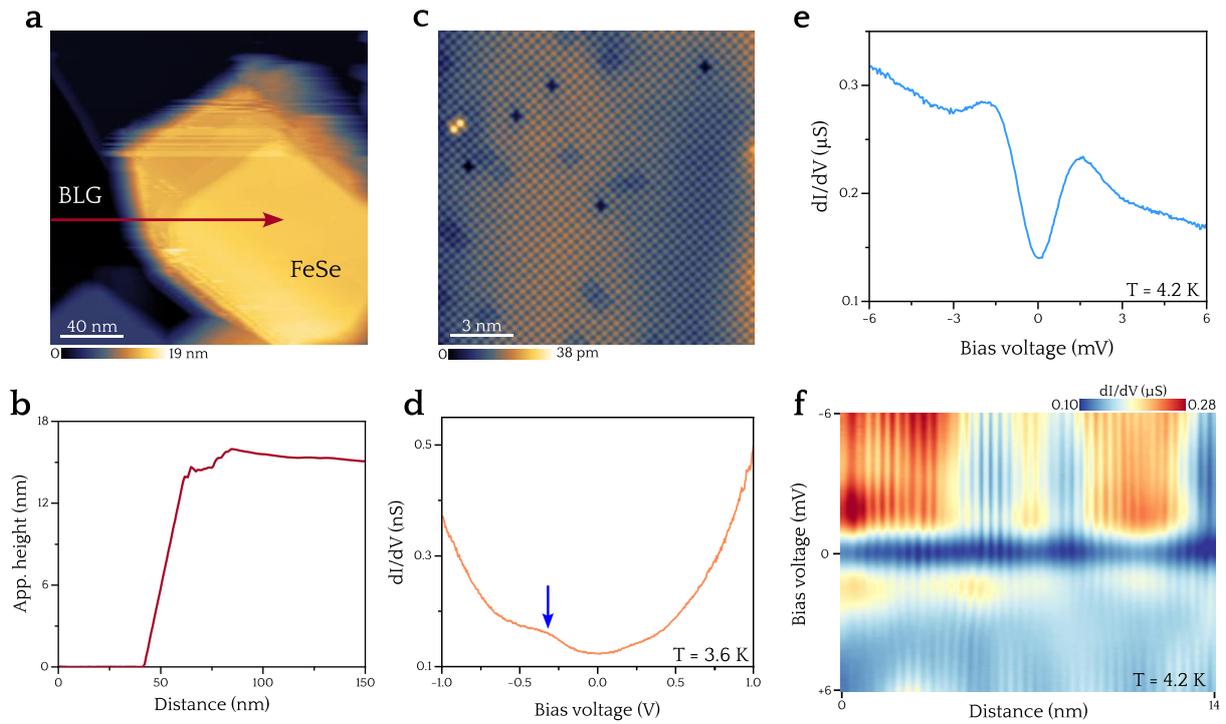

**Figure S4. Electronic and superconducting properties of multilayer FeSe.** a) STM topography image of FeSe/BLG/SiC(0001) surfaces grown at lower temperature (350 ºC) by MBE. Regions of BLG and FeSe are highlighted. b) Apparent height profile measured along the red arrow indicated in (a), highlighting a layer thickness of ≈ 45 layers of FeSe. c) Representative atomically resolved STM image recorded on the surface of the 3D island of FeSe. d) Large-scale electronic structure of multilayer FeSe. e) High resolution d$I$/d$V$ curve showing the development of a superconducting coherent gap in multilayer FeSe. f) Spatially resolved d$I$/d$V$ spectra mapping of the superconducting tunneling gap of thin FeSe films. Acquisition parameters: a) $V_{set}$ = 1 V, $I_{set}$ = 15 pA, $T$ = 3.6 K. c) $V_{set}$ = 0.2 V, $I_{set}$ = 60 pA, $T$ = 3.6 K. d) $V_{mod}$ = 5 mV, $T$ = 3.6 K. e) $V_{mod}$ = 50 µV, $T$ = 4.2 K. f) $V_{mod}$ = 50 µV, $T$ = 4.2 K.

**Supplementary note 5: Electronic structure near Fermi level in multilayer FeSe**

For comparison with few-layer FeSe, Figure S5 presents the electronic structure of multilayer FeSe near the Fermi level, measured across different energy windows. In each case, the differential conductance spectra clearly reveal the presence of a robust, coherent superconducting gap.

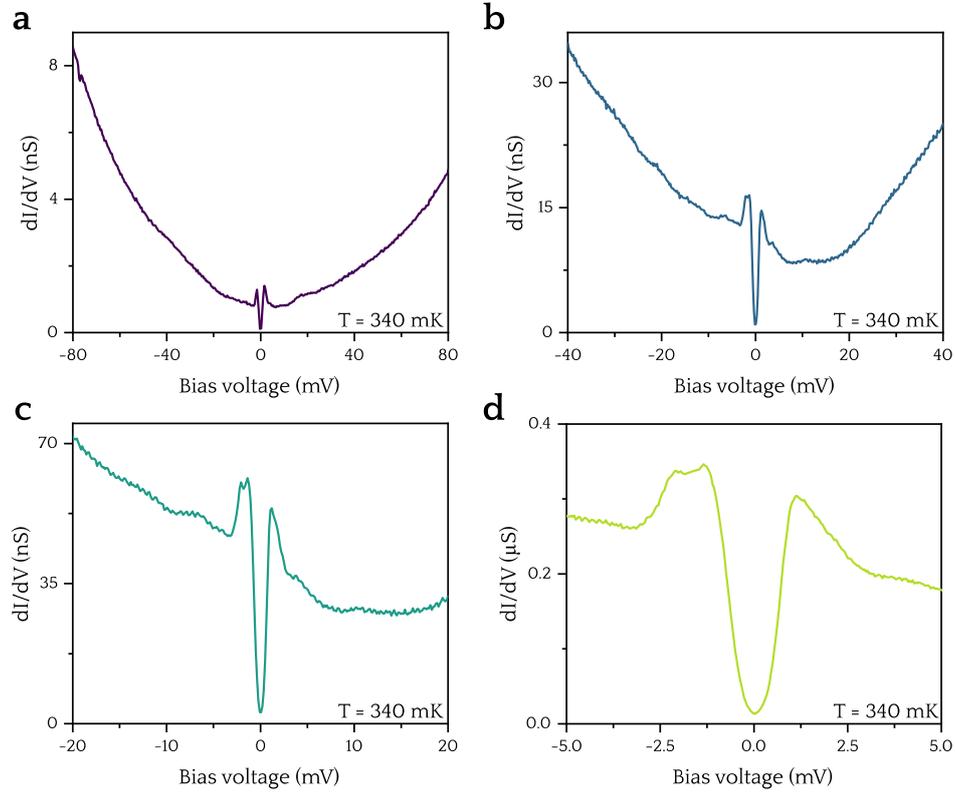

**Figure S5. Electronic structure near Fermi level in multilayer FeSe.** Set of differential conductance spectra measured at different energy scale accounting for the electronic structure of multilayer FeSe near the Fermi level. Acquisition parameters: a) $V_{mod} = 450$ μV, $T = 340$ mK. b) $V_{mod} = 200$ μV, $T = 340$ mK. c) $V_{mod} = 150$ μV, $T = 340$ mK. d) $V_{mod} = 50$ μV, $T = 340$ mK.

**Supplementary note 6: Temperature dependence evolution of the superconducting gap in multilayer FeSe**

To estimate $T_C$ in multilayer FeSe, we studied the evolution of the quasiparticle tunneling spectra with temperature. Figure S6 shows this evolution, where the superconducting gap is gradually suppressed as $T$ increases. We estimate a value of 6 K for the critical temperature, slightly lower than the $T_C$ of bulk FeSe (8 K). This discrepancy can be attributed to the reduced thickness of our multilayer FeSe grown on BLG/SiC(0001), which likely still not fully exhibit bulk-like properties.

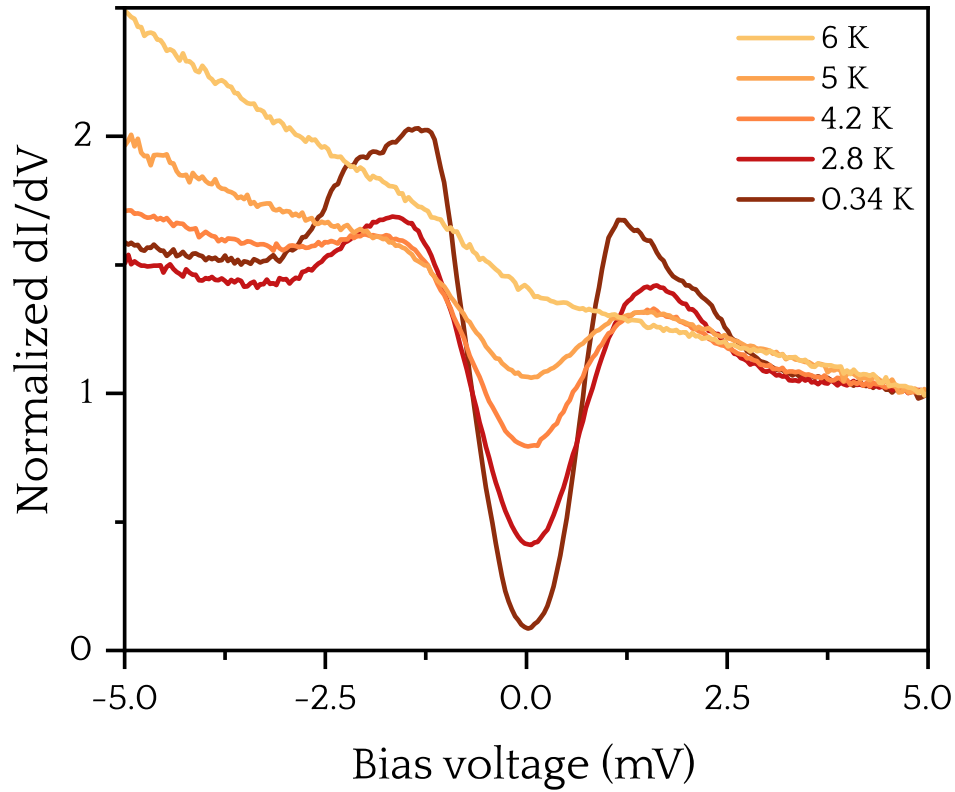

**Figure S6. Temperature dependence evolution of the superconducting gap in multilayer FeSe.** a) Sequential d$I$/d$V$ curves recorded on multilayer FeSe showing the evolution of the superconducting gap as function of the temperature. Acquisition parameters for all curves: $V_{mod}$ = 50 µV.